\documentclass[prapplied, amsmath,amssymbn,superscriptaddress]{revtex4-1}
\usepackage{graphicx}
\usepackage{dcolumn}
\usepackage{bm}

\usepackage{mathptmx}
\usepackage{etoolbox}

\usepackage{setspace}
\usepackage[utf8]{inputenc} 
\usepackage[T1]{fontenc}    
\usepackage{amsfonts}       
\usepackage{amssymb}
\usepackage{amsmath}
\usepackage{booktabs}       
\usepackage{hyperref}       
\usepackage{url}            
\usepackage{microtype}      
\usepackage{nicefrac}       
\usepackage{paralist}       
\usepackage{graphicx}
\usepackage{float}
\usepackage{wrapfig}
\usepackage[dvipsnames]{xcolor}
\usepackage{ulem}
\usepackage[version=3]{mhchem} 

\usepackage{lmodern}
\usepackage{ulem}

\usepackage{subcaption}

\begin{document}

\title{Benchmarking machine-learning interatomic potentials using classical force fields}
\author{Johannes Sandberg}
\email[\textit{Corresponding Author:} ]{johannes.sandberg@univ-lille.fr}
\affiliation{Université de Lille, CNRS, Unit\'{e} Mat\'{e}riaux et Transformations, Lille, France}

\author{Genevi\`eve Dusson}
\affiliation{Université Marie et Louis Pasteur, CNRS, LmB (UMR 6623), F-25000 Besançon, France}

\author{Julien Lam}
\affiliation{Université de Lille, CNRS, Unit\'{e} Mat\'{e}riaux et Transformations, Lille, France}

\begin{abstract}
    We propose a rational comparison between several well-known machine-learning interatomic potentials (linear atomic cluster expansion, the Behler--Parrinello high-dimensional neural network potential, and the graph neural network-based MACE potential) by testing their performance for fitting classical potentials such as the Lennard--Jones and Stillinger--Weber potential. 
    With this we assess their advantages and limitations regarding several aspects: varying training dataset size, robustness to noise, carbon footprint,  as well as the ability to reproduce the radial distribution function as an example observable.
    Overall the MACE models are shown to most accurately reproduce the potential energy surface, with little sign of overfitting, but at a significant computational cost, highlighting important questions regarding the choice of models in computational materials science.
\end{abstract}
\maketitle

\section{Introduction}
Atomistic modeling of materials has been pivotal in the recent development of materials science~\cite{Voter2002Aug,Carter2008Aug}.
The rapid improvements in hardware and emergence of large scale computing  has significantly improved the prediction of novel material properties, and enabled understanding of fundamental processes otherwise unreachable through experimental measurement.
Regardless of the materials studied, such numerical modeling relies on the ability to compute energies and forces among the constituent atoms.
First principles calculations solve the underlying quantum-mechanical electronic structure problem using techniques such as Hartree--Fock~\cite{Echenique2007Dec}, post-Hartree--Fock~\cite{Magnasco2013Jan}, Density Functional Theory~(DFT)~\cite{Cohen2008Aug,Mardirossian2017Oct} and tight-binding methods~\cite{Seifert2007Jul,Seifert2012May}.
This has some strong advantages, allowing for (i) very accurate calculations that can sometimes exceed experimental precision, and (ii) very high flexibility since virtually any material can be modeled.
However, this accuracy and flexibility comes with a tradeoff, as the computational demands of \textit{ab initio} methods prevent from running large-scale simulations that are crucial to capture numerous physical and chemical processes such as phase transitions and nanoparticle structure.
Empirical interatomic potentials, including Lennard--Jones~(LJ) for noble gases, Morse for diatomic molecules, and Embedded Atom Models (EAM) for metals~\cite{Stone2008Aug,Choudhary2017Jan,Harrison2018Sep}, have offered an alternative to \textit{ab initio} approaches.
Such potentials have long formed the backbone of atomistic modeling, and overcome the large computational cost of \textit{ab initio} calculations, allowing for large scale simulations of billions of atoms over long timescales.
While they are computationally efficient, the simple nature of empirical potentials limits their use to qualitative studies, as they are generally not able to provide quantitatively accurate results.
Beyond this limitation, the development of empirical potentials requires chemical and physical intuition, in order to propose suitable functional forms and optimize their parameters.

It is in this context that Machine-Learned Interatomic Potentials (MLIPs) were developed, to enable (1) large scale simulations with (2) the accuracy of \textit{ab initio} methods and (3) a sufficient flexibility that it can, in principle, be fitted for any given material~\cite{Rupp2015Aug,Behler2016Nov,Schleder2019May,Schmidt2019Aug}.The generic idea is to start with a very universal formulation for the MLIP and then use a database made of atomic configurations with associated forces and energies to fit the potential parameters.
Many MLIP approaches have been proposed, including high-dimensional neural network potentials~(HDNNP)~\cite{Behler2017Oct}, Gaussian approximation potentials~\cite{Bartok2010Apr}, the spectral neighbor analysis potentials~\cite{Thompson2015Mar,Wood2018Jun}, linearized potentials~\cite{Seko2015Aug,Mizukami2014Oct,Takahashi2017Nov,Takahashi2018Jun}, the Atomic Cluster expansion~(ACE)~\cite{drautz2019atomic,dusson2022atomic}, and more recent equivariant graph neural networks~(GNN)~\cite{batzner20223,batatia2022mace}.
All of these methods have allowed for a substantial progress in materials modeling, and accurate MLIPs have already been obtained for numerous materials including monoatomic metals~\cite{Deringer2018Apr,Seko2015Aug,Bartok2018Dec}, carbonous materials~\cite{Deringer2018Nov,Deringer2017Mar,Caro2018Apr}, metallic oxides~\cite{Artrith2016Mar,Artrith2011Apr}, water~\cite{Handley2009Jun,Morawietz2016Jul,Nguyen2018Jun} and ternary materials~\cite{Li2017Dec,Mocanu2018Sep}.
Numerical implementations of these methods are being provided and can be employed almost automatically~\cite{Bartok2015Aug,Artrith2016Mar,Thompson2015Mar,singraber2019library,Khorshidi2016Oct,Kolb2017Apr,Yao2018Feb,Wang2018Jul}.

With the multitude of methods available, and rapid development of new ones, with increasing complexity, it is crucial to provide benchmarking that can exhibit the advantages and drawbacks of different approaches.
In this work we provide an alternative to benchmarking on expensive DFT data, by instead evaluating models using classical potentials.
We construct datasets for three reference models of differing complexity: (1) Lennard--Jones~(LJ), (2) an Embedded-Atom Model~(EAM), and (3) a Stillinger--Weber~(SW) model.
For these datasets we train models of increasing complexity, linear models based on ACE~\cite{drautz2019atomic}, Behler--Parrinello HDNNPs~\cite{behler2007generalized}, and MACE graph-neural-network models~\cite{batatia2022mace}.
We examine the ability of each model to learn the forces of the classical potentials, as well as the ability to reconstruct exact pair-potential curves, and four-body interactions, making use of the exactly known reference models.
Simulations are performed to test the ability to reconstruct the radial distribution of the liquid.
The sensitivity to noise and dataset size is tested as well.
Finally we provide an analysis of the computational cost of running simulations, and the corresponding carbon footprints of the models are estimated and compared, to our knowledge for the first time for MLIPs.
We show that the MACE model is in many ways superior in its ability to the other models, but at the price of a severely increased computational cost, and higher carbon footprint, with the ACE models still achieving very acceptable results in most cases, at a way lower computational cost, highlighting computational tradeoffs appearing for very large systems.

\section{Method}
\subsection{Database Construction}

\begin{table}
    \centering
    \begin{tabular}{lc|cc}
        & System & $T_0$ & $T_1$  \\\hline
        Liquid 1 & EAM & $2000$~K & $3000$~K\\
        & LJ & $71$~K & $82$~K\\
        & SW & $2000$~K & $3500$~K\\\hline
        Liquid 2 & EAM & $2000$~K & $5000$~K\\
        & LJ & $71$~K & $106$~K\\
        & SW & $2000$~K & $7000$~K\\\hline
        Cluster & EAM & $2000$~K & $2000$~K\\
        & LJ & $71$~K & $71$~K\\
        & SW & $2000$~K & $2000$~K\\
    \end{tabular}
    \caption{Parameters of the datasets used in this work.}
    \label{tab:dataset_params}
\end{table}

In this work we consider three reference models: a LJ potential, parameterized for \ce{Ar}, an EAM parameterized for \ce{Cu}~\cite{foiles1985calculation}, and SW parameterized for \ce{Si}~\cite{stillinger1985computer}.
For LJ the parameters are $\epsilon = 117.7$~K$/k_\mathrm{B}$, $\sigma=3.504$~\AA, with the potential truncated at $r_c=2.5\sigma$ and shifted by subtracting the value of the pair potential at this radius.
For each potential we construct three training databases through sampling molecular dynamics trajectories obtained through simulation in LAMMPS~\cite{thompson2022lammps}.
In all simulations a timestep of $1$~fs was used, a Nose--Hoover thermostat~\cite{nose1984molecular} was used to control temperature, and a barostat ensured ambient pressure.
For the first dataset (Liquid 1) a system of 64 atoms are placed randomly in a simulation box with a density of $1$~atom$/$\AA$^3$, and an energy relaxation is performed to obtain an initial configuration.
Initial velocities were then assigned via a Boltzmann distribution at temperature $T_0$, and a simulation with $10^6$ timesteps was performed to equilibrate the system at fixed temperature $T_0$.
Sampling of the trajectory was then performed during a $10^6$ timestep simulation with a temperature ramp from $T_0$ to $T_1$, extracting $250$ snapshots at regular intervals of $4000$ timesteps.
A second dataset (Liquid 2) was constructed in an identical manner, but with a larger final temperature $T_1$, chosen such that the system starts to evaporate at the end of the simulation, ensuring sampling of much less dense configurations.

Besides the two liquid datasets, we construct a third dataset (Cluster) consisting of snapshots of small 64-atom clusters.
The simulations were performed similarly to the liquid, apart from the initialization wherein a spherical cluster of $64$ atoms is constructed by placing atoms at random positions inside a sphere, as opposed to the entire simulation box.
In order to avoid evaporation of the droplets, a Lennard--Jones interaction was applied between the atoms and the walls of the simulation box, with the energies later recomputed without this boundary interaction.
The parameters for all the database simulations are given in table~\ref{tab:dataset_params}.

\subsection{Machine Learning Interaction Potentials}
A variety of MLIP models can be found in the literature, employing different representations of the atomic environment, and different machine learning regression schemes.
In this work we compare three established MLIP methods: the linear ACE, HDNNP, and MACE.

\textit{Linear ACE:}
The Atomic Cluster Expansion (ACE)~\cite{drautz2019atomic,dusson2022atomic} is a method for constructing a complete set of basis functions for atomic properties.
Such a basis set can then directly be used to construct a linear model of the interaction potential, by expanding a local atomic energy in this ACE basis:
\begin{equation}
    E_i = \sum_k c_k B_{i,k},
\end{equation}
and has been employed to construct potentials for a wide variety of systems~\cite{lysogorskiy2021performant,ibrahim2023atomic,qamar2023atomic,pan2024atomic,rosset2026atomic}.
The ACE basis is constructed through symmetrized products of spherical harmonics and radial basis functions, as detailed in other works~\cite{drautz2019atomic,dusson2022atomic}.
Here we use spherical Bessel functions for the radial basis, defined in \cite{bochkarev2022efficient}.
We use a maximum body order of $\nu=6$ with maximum radial expansion per body order $n_{\mathrm{max}, \nu} = (32, 6, 4, 3, 2, 1)$.
The angular expansion per body order is taken as $l_{\mathrm{max}, \nu} = (0, 3, 3, 2, 2, 1)$.
We perform all ACE calculations using the \texttt{python-ace} package~\cite{lysogorskiy2021performant, bochkarev2022efficient}, with linear fitting being done using ridge regression.
The ACE basis settings were taken from example settings on the \texttt{python-ace} git repository~\cite{aceExampleCuII}, but with an increased $\nu=0$ radial cutoff.

\textit{Behler--Parrinello HDNNP:}
The Behler--Parrinello HDNNP~\cite{behler2007generalized} constructs the PES as a sum of local atomic energies, fitted as feed-forward neural networks applied to rotationally and translationally invariant descriptors capturing the local environment of each atom.
Here we use the conventional radial and angular Symmetry Functions (SF) with a cosine cutoff~\cite{behler2011atom}.
For each element we generate SF parameters using an automatic enumeration scheme as described in the literature~\cite{gastegger2018wacsf,imbalzano2018automatic}.
Each atomic Neural Network Potential (NNP) consisted of two hidden layers of $10$ neurons each, with $10$ radial and $12$ angular SFs, with parameters given in the supplementary information.
All HDNNP training and calculations were performed using the Neural Network Potential Package (N2P2)\cite{singraber2019parallel}.

\textit{MACE:}
The message passing ACE (MACE) potential is a equivariant graph neural network potential model, with many-body messages constructed using an ACE basis~\cite{batatia2022mace}.
A node of the graph neural network, at layer $t$, encodes the space coordinates $\mathbf{r}_i$, species $z_i$, and a learnable feature vector $\mathbf{h}_i^t$, into the label $\sigma_i^t=(\mathbf{r}_i,\sigma_i,\mathbf{h}_i^t)$.
At each layer, messages $m_{i}$ are constructed as a linear expansion in the ACE basis of atom $i$.
The features at the subsequent layer are then updated as a linear combination of the messages of adjacent nodes, with a residual contribution.
Ultimately, the atomic energy of atom $i$ is constructed a sum of readouts from the invariant features of each layer.
For the final layer the readout consists of a trainable multi-layer perceptron, with earlier layers having a linear readout.
In-depth details about the MACE architecture can be found in \cite{batatia2022mace,kovacs2023evaluation}.
All our MACE models are implemented using the code from~\cite{batatia2022mace}, with recommended parameters of $128$ equivariant messages with an angular resolution of $l_\mathrm{max}=3$.

\section{Results}
\subsection{Comparison between models}

\begin{figure*}
    \centering
    \includegraphics[scale=0.6]{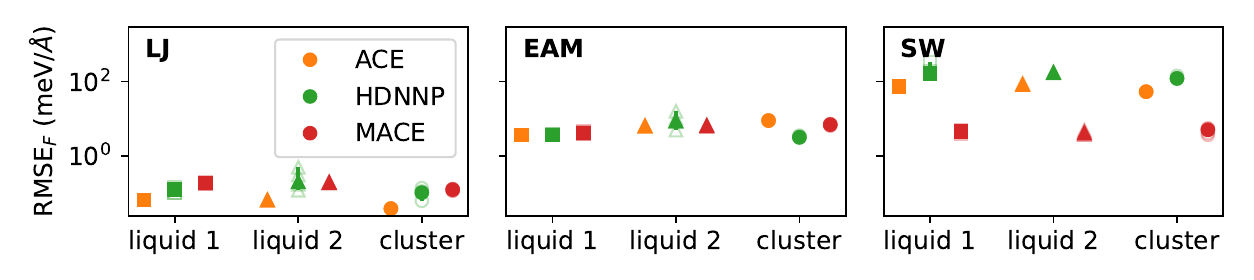}
    \caption{Force RMSE of the different models, for different systems and datasets. Unfilled symbols show values for models trained on different dataset splits, vertical lines indicating the range of values, and solid symbols showing median values.}
    \label{fig:main_run_F}
\end{figure*}

For each system (LJ, EAM, SW), and for each dataset (Liquid 1, Liquid 2, Cluster), we train a set of models (linear ACE, HDNNP, MACE).
Each model is trained five times, on each dataset, with different random partitions of the dataset into training and test sets at a ratio of 9 to 1.
A first measure of the quality of a model is the Root Mean Square Error (RMSE) evaluated on the forces.
This, along with the RMSE on the per atom energies, is ultimately what is optimized during training.
We present in Figure~\ref{fig:main_run_F} the RMSE of all models on the different systems, grouped by dataset.
For each system, the force RMSEs are evaluated on the same held-out test dataset, comprising $250$ structures generated in an identical way to the training dataset, but with a different initial configuration.
We observe that the error typically decreases as we go from the Liquid 1 dataset, to the Liquid 2 dataset, to the Cluster dataset, although with fairly small differences, which is expected since the lower density should allow the few-body nature of the LJ and SW models to be more transparent in the dataset.
For LJ we observe the linear ACE model performing the best, followed by the HDNNP, and the MACE.
Nevertheless, the force error, compared to what is typical in a DFT-based setting, is exceptionally low for all models.
For the EAM system we observe, again, an exceptionally low error, with the models performing roughly equivalently, apart from the HDNNP which has a lower RMSE for the cluster dataset, compared to the other models. As this stage, it remains that the errors are so small that the difference between models should probably not be considered as significant. Meanwhile, more drastic differences between the models are observed in the SW system, where neither the linear ACE or the HDNNP manage to compete with MACE, with over an order of magnitude difference, the latter being able to reach equivalent error as for the EAM system.

\begin{figure}
    \centering
    \includegraphics[scale=0.28]{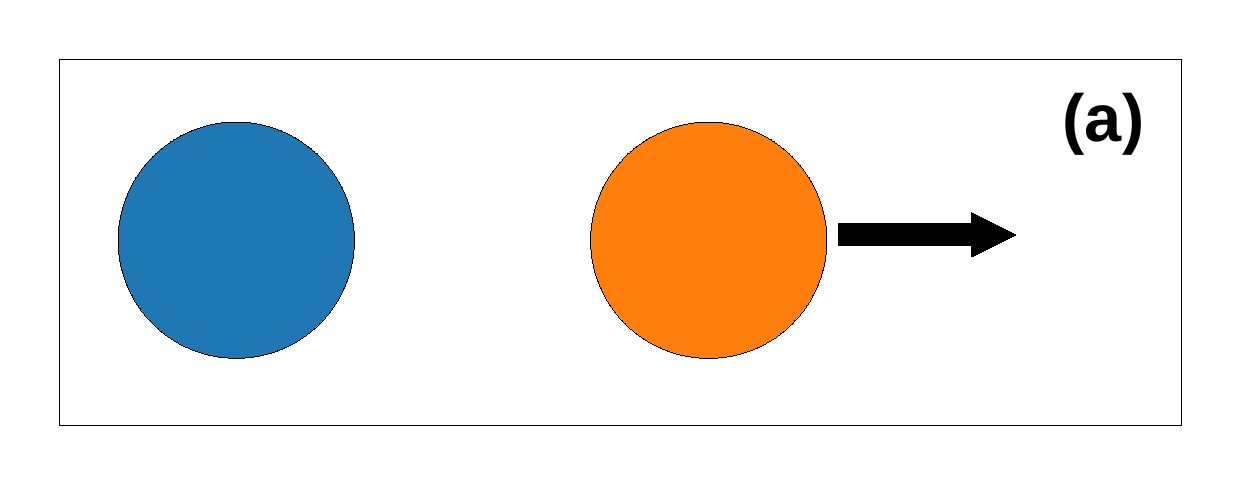}\\
    \includegraphics[scale=0.28]{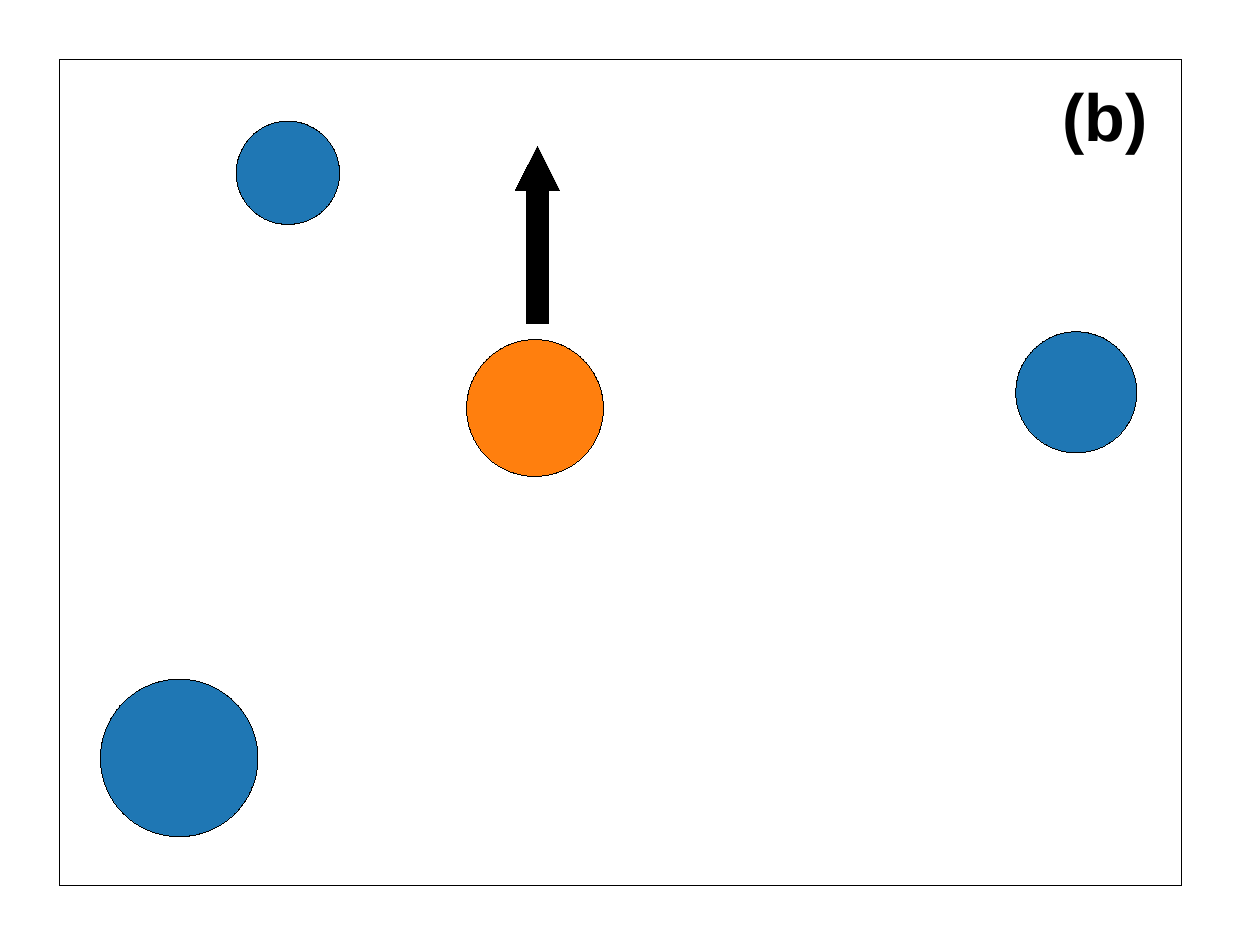}
    \caption{Illustration of the systems used to test the two-body (a) and four-body (b) interaction. Stationary atoms are shown in blue, with probe atoms shown in orange, with arrows indicating the movement of the probe atom.}
    \label{fig:2B_4B_diagram}
\end{figure}

\begin{figure*}
    \centering
    \includegraphics[scale=0.6]{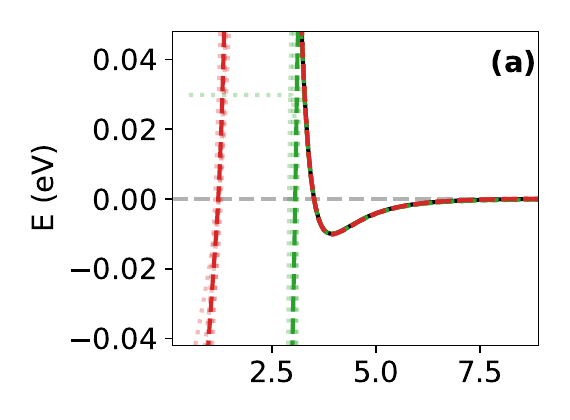}
    \includegraphics[scale=0.6]{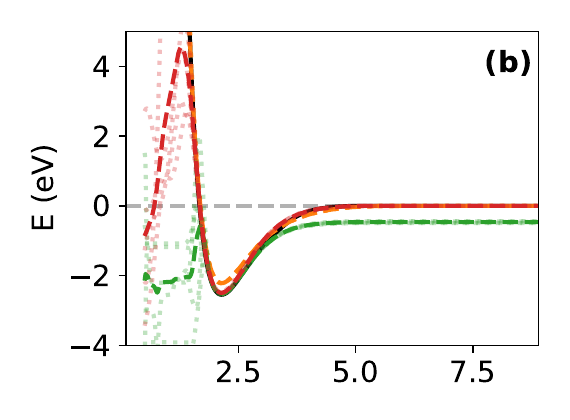}
    \includegraphics[scale=0.6]{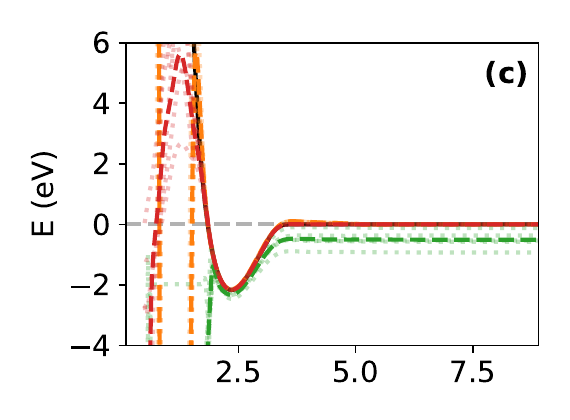}
   \\ 
    \includegraphics[scale=0.6]{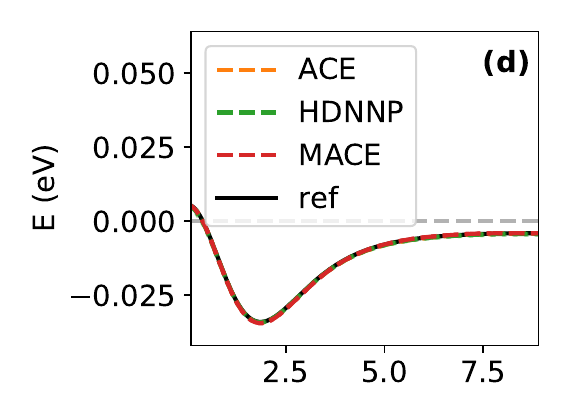}
    \includegraphics[scale=0.6]{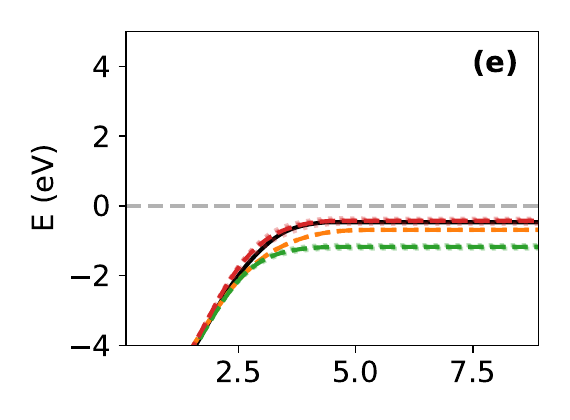}
    \includegraphics[scale=0.6]{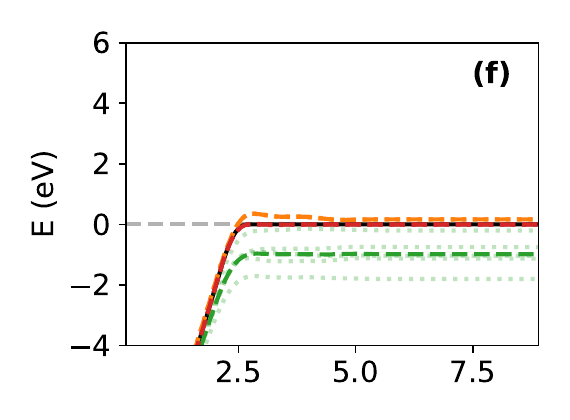}
    \caption{2-body potential energy curves for LJ (a), EAM (b), and SW (c), and the 4-body potentials energy curve for LJ (d), EAM (e), and SW (f).
    Black lines show the underlying potential, transparent dotted lines show curves for models trained on different splits of the cluster dataset, with dashed lines showing averages.}
    \label{fig:2B_4B_single}
\end{figure*}

\begin{figure*}
    \centering
    \includegraphics[scale=0.6]{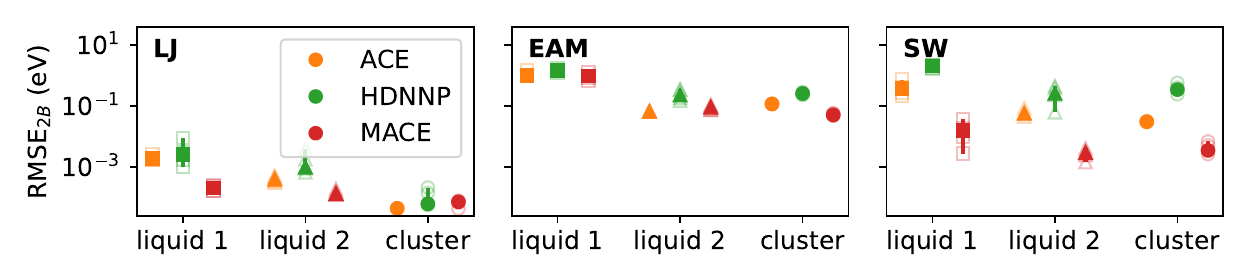}
    \includegraphics[scale=0.6]{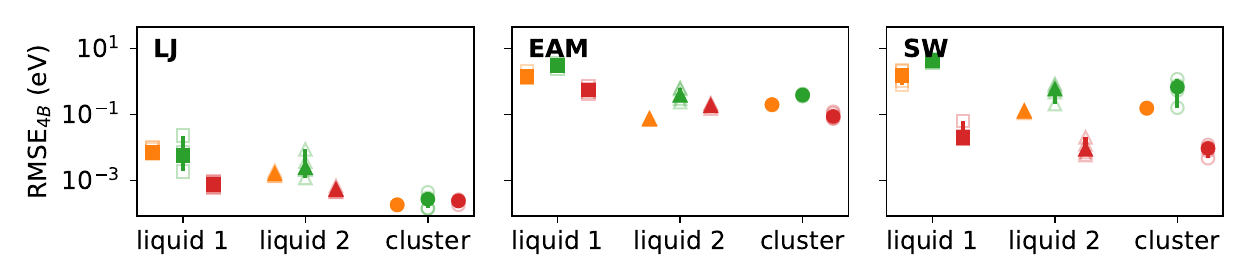}
    \caption{RMSE in the 2-body potential-energy (top), 4-body potential energy (bottom), for different models, on different systems and datasets.}
    \label{fig:main_run}
\end{figure*}

In order to study the ability of the models to reproduce the underlying classical potential, we consider two test cases illustrated in Figure~\ref{fig:2B_4B_diagram}.
The first test, Figure~\ref{fig:2B_4B_diagram}(a), consists of a simple dimer system, with one atom placed at the origin, and a probe atom placed at a distance of $0.5~\AA$ from the central atom.
The probe atom is moved in a positive radial direction, and the potential energy recorded, in order to extract the pair interaction.
A second test case consists of three atoms, placed in a equilateral triangle configuration, as illustrated in Figure~\ref{fig:2B_4B_diagram}(b), with a probe atom placed in the center of the configuration.
The probe atom is moved in the normal direction to the triangle, and the potential energy recorded, in order to study the 4-body interaction.
Figure~\ref{fig:2B_4B_single} shows the 2-body potential energy curves for the LJ (a), EAM (b), and SW (c) models trained on the cluster dataset, as well as the 4-body potential energy curve for the same (d-f).
We see that the MACE is able to reproduce the underlying PES impressively well in all three systems.
The linear ACE is less able to reproduce the exact PES, except for LJ, underestimating the potential well for EAM, and erroneously predicting a small hump just outside the energy well in SW.
We observe that the HDNNP manages to reproduce fairly well the bottom of the potential well, but has failed to learn the short-range repulsion, indicating instability in the potentials.
For both tests cases we calculate the RMSE between the curve for the reference model, and that for the ML models showing the result in Figure~\ref{fig:main_run}.
Here we observe more drastic differences between the models, compared to the force RMSE.
While the differences between different datasets were fairly modest, in terms of improving the prediction of forces, we here observe a clear trend, with the Cluster dataset having lower error compared to Liquid 1 and 2, and Liquid 2 having a lower error than Liquid 1.
When comparing the three models, the MACE is generally much better at reproducing the underlying potential, which is surprising considering the highly complex architecture of MACE, compared to the HDNNP and linear ACE potential.

\begin{figure}
    \centering
    \includegraphics[scale=0.6]{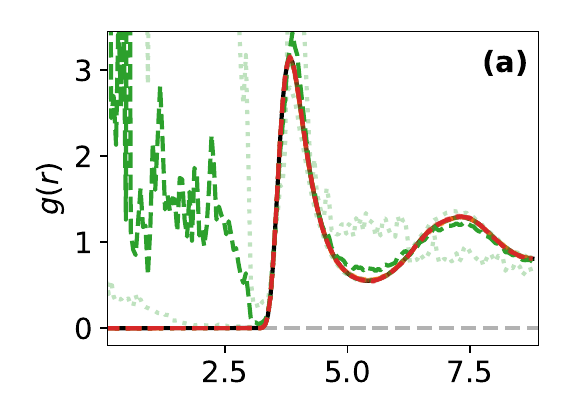}
    \includegraphics[scale=0.6]{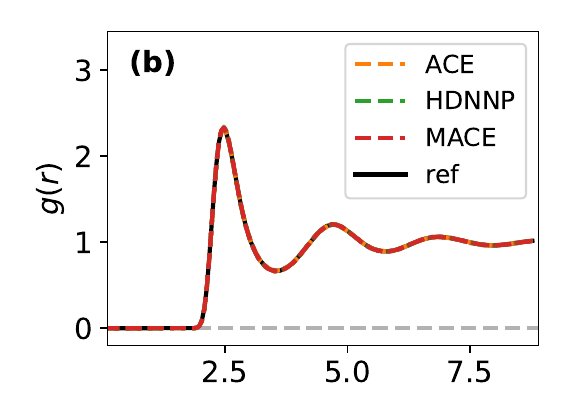}
    \includegraphics[scale=0.6]{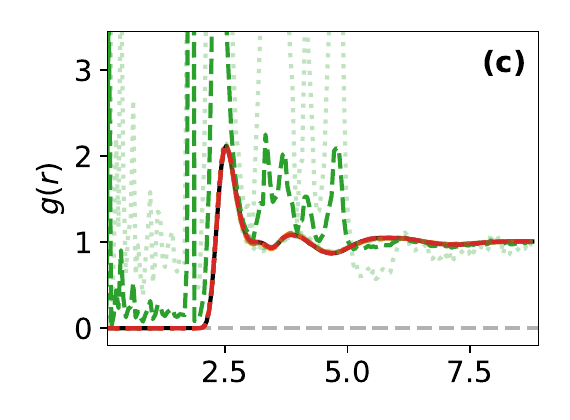}
    \caption{Exact and predicted RDFs for LJ (a), EAM (b), and SW (c), with models trained on the cluster dataset. Transparent dotted lines show curves for models trained on different dataset splits, with dashed lines showing averages.}
    \label{fig:rdf_single}
\end{figure}

\begin{figure*}
    \centering
    \includegraphics[scale=0.6]{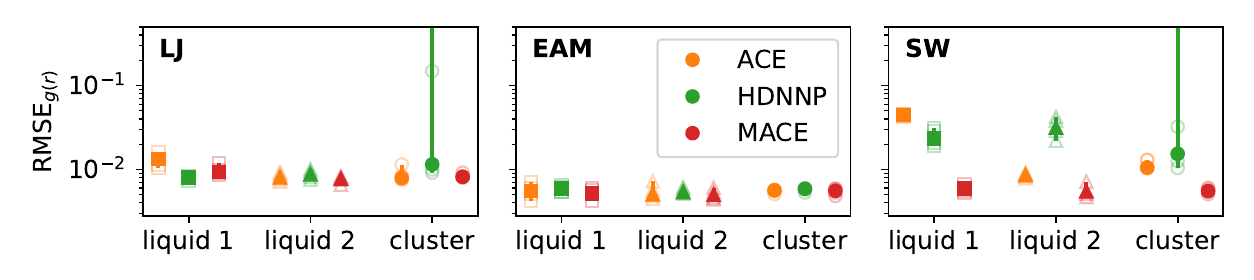}
    \caption{RMSE in the radial distribution function, for different models, on different systems and datasets as in Figure~\ref{fig:main_run_F}.}
    \label{fig:main_run_rdf}
\end{figure*}

Finally, we perform simulations of a bulk liquid, with the aim of extracting $g(r)$, the radial distribution function (RDF).
In each system we prepare a box of $512$ atoms, placed randomly followed by an energy minimization with the classical potential.
For the EAM and SW system we perform an NVT simulation at $4000$~K, over $100$~ps, followed by a $200$~ps NPT simulation at a pressure of $1$~bar, and a temperature of $2000$~K.
The same process was followed for LJ, but at temperatures of $141$~K and $71$~K.
At the end of the second simulation, the final configuration is saved, for use as a starting point in subsequent simulations.
For each ML potential, and the reference, the system is initialized from this starting configuration, and simulation is continued for $200$~ps, over which the RDF is calculated.
Figure~\ref{fig:rdf_single} shows the predicted RDFs for LJ (a), EAM (b), and SW (c), with the models trained on the cluster dataset.
We see a perfect agreement with the reference value for EAM, across all models.
For LJ and SW, we see a good agreement for MACE, and for the linear ACE, but very clear signs of instability for the HDNNP model.
Figure~\ref{fig:main_run_rdf} shows the RMSE between the RDF calculated for the reference systems, and each MLIP model.
Apart from the SW system, we observe only relatively small differences between the different models in the predicted RDFs, despite the differences in the accuracy of predictions of the underlying PES.
It should be noted that some of the HDNNP potentials were unstable, visible in Figure~\ref{fig:main_run}.

In Figures~\ref{fig:main_run_F} and \ref{fig:main_run} it can be seen that some of the models have a wider range of errors, across the different dataset splits.
This is further illustrated by the standard deviations of the RMSE for the forces, pair interaction, and RDF, shown in the SI.
We observe that the linear ACE typically has the most consistent force error, explainable by the non-stochastic nature of the training method, as well as the use of regularization.

\subsection{Influence of dataset size and noise}

\begin{figure}
    \centering
    \includegraphics[scale=0.6]{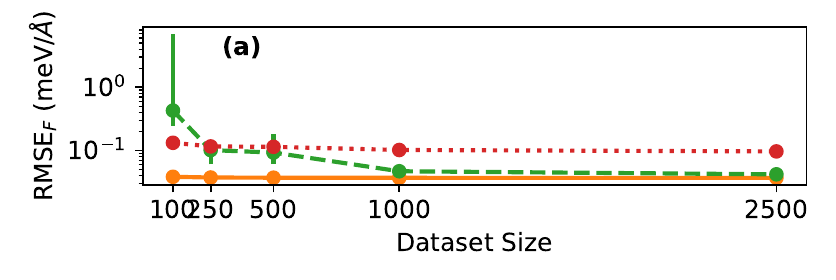}
    \includegraphics[scale=0.6]{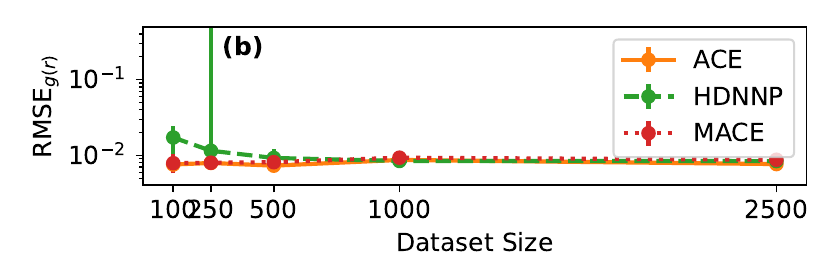}
    \includegraphics[scale=0.6]{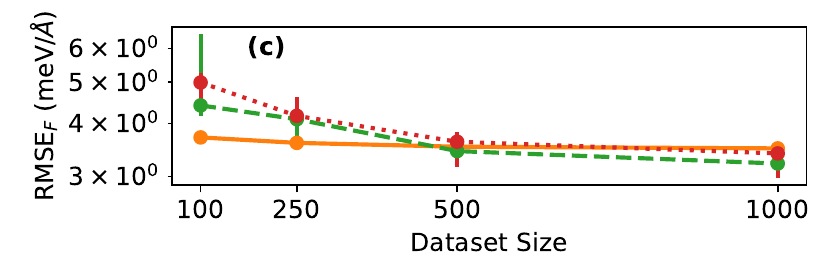}
    \includegraphics[scale=0.6]{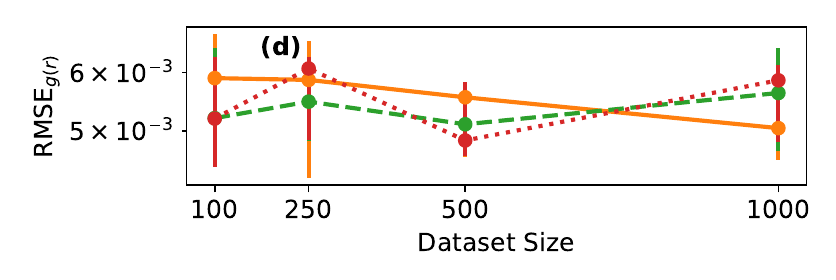}
    \caption{Influence of the dataset size.
        (a) The RMSE in the forces, and RDF (b), as function of dataset size for the LJ cluster dataset.
        (c) The RMSE in the forces, and RDF (d), as function of dataset size for the EAM Liquid 1 dataset.        (c) The RMSE in the forces, and RDF (d), as function of dataset size for the EAM Liquid 1 dataset.
        Markers show averages across models trained on different dataset splits, with vertical bars showing standard deviation.} 
    \label{fig:dataset_size}
\end{figure}

We now examine the influence of the training dataset size on the models.
To this end we focus on the EAM Liquid 1 dataset and the LJ cluster dataset, as examples for which we got good results in the previous section.
For each we create a sequence of datasets of varying size, by extending the length of the underlying trajectories, keeping the sampling rate constant.
For each of these we train and test a set of models as previously.
Figure~\ref{fig:dataset_size}(a)-(b) shows the dependence of the RMSE in the forces, and RDF, respectively for LJ cluster, as function of the dataset size, including also the original dataset size of $250$ structures.
We see a force error that is fairly constant for the MACE and ACE, but has a much stronger dependence on the dataset size for the HDNNP.
Figure~\ref{fig:dataset_size}(c)-(d) shows the RDF and force RMSE for the EAM liquid, as function of dataset size.
Here we observe a small improvement in the force error, as the dataset size increases, with the linear ACE having the lowest error for the smallest datasets, but also least improvement with additional data, as the error of the different models converge around $500$ training structures.
No significant dependence on the dataset size is observed in the error on the RDF.

\begin{figure}
    \centering
    \includegraphics[scale=0.6]{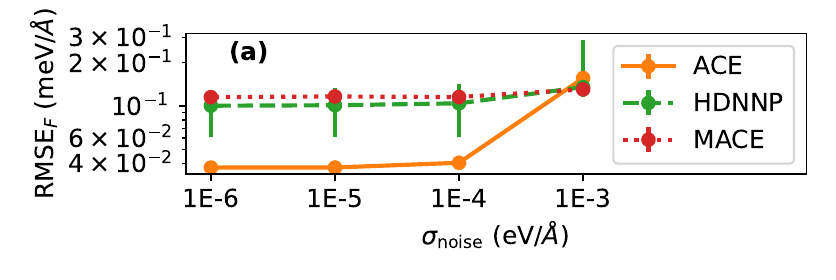}
    \includegraphics[scale=0.6]{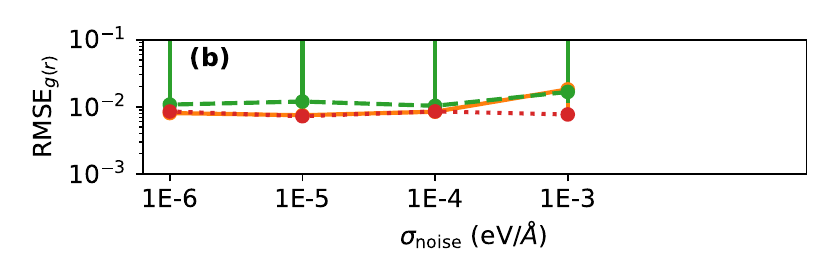}
    \includegraphics[scale=0.6]{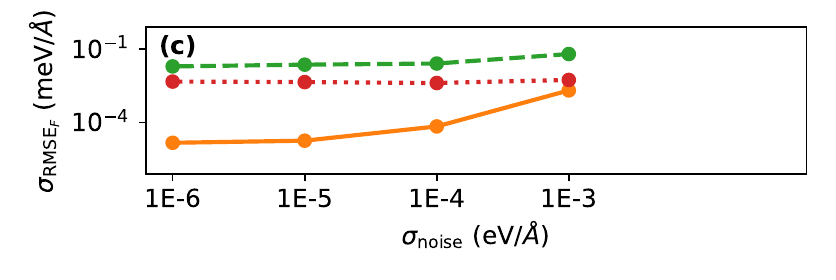}
    \caption{Influence of noise in the dataset.
        (a) The RMSE in the forces, with respect to noiseless data, and RDF (b), for models trained on noisy data as function of the scale of noise added to the force labels.
        (c) Standard deviation of the data in (a).}
    \label{fig:noise}
\end{figure}

In order to test the sensitivity of the models to the presence of noise in the training data, we take the LJ cluster dataset, being the simplest system and thus a good choice for isolating the effects of noise, and modify it by adding a Gaussian noise, with varying standard deviation.
The noise is added to the force and energy values.
Inspection of the dataset showed that the standard deviation of the energy was approximately $4$ times larger that for the forces, in the unmodified dataset, and as such the noise added to the energies was increased by a factor of $4$.
Figure~\ref{fig:noise}~(a-b) shows the RMSE of the forces and RDF, for all three models, as a function of the scale of noise added to the forces, tested on the noise-free data.
As in Figure~\ref{fig:main_run_F}, the ACE model achieves the lowest error, followed by HDNNP, and MACE.
In all cases we observe a fairly constant error, with the noise having little impact at $10^{-1}$~$meV/\AA$ and below.
Above this scale, there is a corresponding increase in the error of the models, coinciding with the scale of the noise exceeding the error on the less noisy data.
Interestingly, the MACE, while having a larger error on this dataset, has by far the least increase in error as the noise level increases.
The ability of the MACE to overcome noise in the dataset is particularly noteworthy, as current state of the art models approach DFT accuracy, and noise in the dataset, for instance from poorly converged calculations \cite{kuryla2025accurate}, become important for the performance of the models.
For the RDF we observe, again, the instability of one of the HDNNPs, which is not affected by the introduction of noise.
Looking at the standard deviation in the force errors, Figure~\ref{fig:noise}~(c), we see that as the noise increases, so does the variance in the force error, with the MACE again seeing the lowest impact from increasing levels of noise.

\begin{table*}
    \centering
\begin{tabular}{l|ccc}
 & ACE & HDNNP & MACE\\\hline
LJ & $0.2664$ s & $0.07355$ s & $61.95$ s\\
EAM & $0.2639$ s & $0.07534$ s & $60.08$ s\\
SW & $0.2707$ s & $0.07478$ s & $60.32$ s\\
\end{tabular}

    \caption{Running times of the 4-body calculations, for models trained on the Liquid 2 dataset.}
    \label{tab:timing_4B}
\end{table*}

\begin{table*}
    \centering
\begin{tabular}{l|ccc}
 & ACE & HDNNP & MACE\\\hline
EAM & 00:03 & 00:14 & 01:05\\
       & $7.11$ Wh ($0.365$ g) & $33.2$ ($1.70$ g) & $672$ Wh ($34.5$ g) \\
LJ & 00:03 & 00:10 & 01:00\\
         & $7.11$ Wh ($0.365$ g) & $23.7$ Wh ($1.22$ g) & $620$ Wh ($31.8$ g) \\
SW & 00:03 & 00:10 & 00:52\\
     & $7.11$ Wh ($0.365$ g) & $23.7$ ($1.22$ g) & $538$ Wh ($27.6$ g)\\
\end{tabular}

    \caption{Running time for the RDF simulation (h:m), for models trained on the Liquid 2 dataset, with estimated energy usage and carbon footprint (g\ce{CO_2}e).}
    \label{tab:timing}
\end{table*}

Beyond the accuracy of MLIPs, and their ability to reproduce physically relevant quantities, the efficiency of the models are also important.
Table \ref{tab:timing_4B} shows the average running time for calculating the 4-body curves of the models trained on Liquid 2 (Figures~\ref{fig:2B_4B_single} and \ref{fig:main_run}), running on a single cpu core.
We observe runtimes below one second for both ACE and HDNNP, with HDNNP being faster for this configuration.
The MACE, in contrast, required over a minute for the same calculation, with more than a factor $200$ difference compared to the ACE.
Table \ref{tab:timing} shows the wall time for running the $200$k timestep simulation for the different models, used to calculate the RDF.
Here, the linear ACE, and HDNNP models all ran on $8$ cpu cores, and can thus be directly compared.
The MACE models each ran on a single gpu, as running on the same set of cpus was in this case not feasible, making a direct comparison to the other models difficult.
Nevertheless, it should be clear from the runtimes that the MACE model is significantly more computationally demanding than the other models.
A more fair metric of model efficiency is the energy consumption of the simulation, which we have estimated following \cite{lannelongue2021green}, with further details given in the SI.
The energy consumption of each model is included in Table~\ref{tab:timing}, along with estimates for the corresponding carbon footprints.
Despite these being rough estimates, the picture is quite clear, with the linear ACE model requiring by far the least energy to run, by about a factor of $5$ compared to the HDNNP, and almost two orders of magnitude compared to the MACE.
It should be mentioned that for this denser system, with more atoms, the ACE was faster than the HDNNP, in contrast to the $4$ body system.

\section{Conclusion}
In this work we have used classical potentials in order to benchmark, and compare, three commonly used MLIP frameworks: linear ACE potentials, Behler--Parrinello HDNNPs, and MACE.
The models have been evaluated in their ability to reproduce reference forces, exact potential energy curves in two test setups, and the radial distribution function as an example observable quantity.
We have studied the effects of dataset size, and of noise, and performed a comparison of computational cost.
We observe a surprising ability of MACE architecture in learning the exact underlying PES, even in extrapolation, despite the huge number of learnable parameters, raising questions about how MACE avoids overfitting.
Furthermore, our results show that MACE is less sensitive to noise in the dataset, compared to the other methods studied here.
This is again a surprising result, considering the enormous number of parameters, as without any form of regularization, one would expect MACE to be in the overparameterized regime, and thus overfit in the presence of noise.
While these results do speak strongly in favor of MACE, the clear downside of the model is the extreme computational cost, potentially limiting its applicability in many situations requiring large systems and long timescales.

All these results, taken together, point to the fact that the choice of MLIP method is not a simple one, with trade-offs between accuracy and computational efficiency, necessitating further research into the merits of each method.
Ultimately, the appropriate choice of method will likely depend on the application, and system under study.
For applications where it is crucial that the PES be perfectly reproduced, regardless of computational cost, or where it will only be calculated infrequently, MACE is clearly the better option out of the models studied here.
For systems where other methods fail, such as the SW system in this work, the versatility of MACE is also a great benefit.
For applications where many potential energy calculations will be performed, such as in crystal nucleation, and rare event sampling more broadly, the lower computational cost of ACE might become much more of a factor.
The situation is further complicated by the ACE often being embedded in a nonlinear function, which could allow better competition with the MACE.
On the other hand, in multicomponent systems there are additional complexities, and the scaling with number of species becomes crucial, with MACE likely a good candidate for systems with a large number of different elements, such as high-entropy alloys.
It is clear that the optimal choice of MLIP model is far from settled.
When comparing different MLIP models, care should also be taken to consider not just the number of timesteps per second achieved in benchmarks, but to also evaluate energy costs and relative environmental impact.

\section*{Acknowledgements}
This work was supported by the CNRS AISSAI Center postdoctoral fellowship program, funded by the CNRS Foundation through a Google donation.
This project was provided with computing HPC
and storage resources by GENCI at CINES/IDRIS/TGCC through grant 2021-A0170913010.
Further computational resources have been provided by the DSI at Université de Lille.

\bibliography{references}

\end{document}